\documentclass[12pt]{article}
\usepackage{amsmath}
\usepackage{amscd}
\usepackage{amssymb}
\usepackage{array}

\begin{document}
\rightline{CERN-TH/2002-073}

\newcommand{\R}{\mathbb{R}}
\newcommand{\C}{\mathbb{C}}
\newcommand{\Z}{\mathbb{Z}}
\newcommand{\Hb}{\mathbb{H}}

\newcommand{\rE}{\mathrm{E}}
\newcommand{\rSp}{\mathrm{Sp}}
\newcommand{\rSO}{\mathrm{SO}}
\newcommand{\rSL}{\mathrm{SL}}
\newcommand{\rSU}{\mathrm{SU}}
\newcommand{\rUSp}{\mathrm{USp}}
\newcommand{\rU}{\mathrm{U}}
\newcommand{\rF}{\mathrm{F}}

\newcommand{\fe}{\mathfrak{e}}
\newcommand{\fsp}{\mathfrak{sp}}
\newcommand{\fusp}{\mathfrak{usp}}
\newcommand{\fsu}{\mathfrak{su}}
\newcommand{\fp}{\mathfrak{p}}
\newcommand{\fso}{\mathfrak{so}}
\newcommand{\fl}{\mathfrak{l}}
\newcommand{\fg}{\mathfrak{g}}
\newcommand{\fr}{\mathfrak{r}}

\newcommand{\id}{\relax{\rm 1\kern-.35em 1}}
\vskip 1.5cm

  \centerline{\LARGE \bf Gauging of Flat Groups in Four}
\bigskip
 \centerline{\LARGE \bf  Dimensional Supergravity}

 \vskip 3cm
\centerline{L. Andrianopoli$^{\sharp\;\natural}$, R.
D'Auria$^{\sharp\;\natural}$, S. Ferrara$^{\sharp\;\flat}$ and M.
A. Lled\'o$^{\sharp\;\natural}$.} \vskip 1.5cm

\centerline{\it $^\sharp$ Dipartimento di Fisica, Politecnico di
Torino,} \centerline{\it Corso Duca degli Abruzzi 24, I-10129
Torino, Italy. } \medskip

\centerline{\it $\natural$  INFN, Sezione di Torino, Italy. }

\medskip

\centerline{\it $^\flat$ CERN, Theory Division, CH 1211 Geneva 23,
Switzerland, and } \centerline{\it INFN, Laboratori Nazionali di
Frascati, Italy.}

\vskip 1cm

\begin{abstract}
We show that $N=8$ spontaneously broken supergravity in four
dimensions obtained by Scherk--Schwarz generalized dimensional
reduction can be obtained from a pure four dimensional perspective
by gauging a suitable electric subgroup of $\rE_{7,7}$. Owing to
the fact that there are non isomorphic  choices of maximal
electric subgroups of the U-duality group their gaugings  give
rise to inequivalent theories. This in particular shows that the
Scherk--Schwarz gaugings do not fall in previous classifications
of possible gauged $N=8$ supergravities. Gauging of flat groups
appear in many examples of string compactifications in presence of
brane fluxes.
\end{abstract}

\vfill\eject

\section{Introduction}
In recent times, investigation of string theory models whose low
energy dynamics is described by a compactification of ten or
eleven dimensional supergravity in presence of brane fluxes, has
revived the interest in the study of gauged supergravities with
different number of supersymmetries, and their possible Higgs and
super Higgs phases \cite{ps} - \cite{kst}. Other examples of
models with broken phases are the Scherk--Schwarz \cite{ss,css}
supergravities and the so called ``no-scale" supergravity models
\cite{cfkn,elnt}. Models of this type naturally appear when brane
fluxes are turned on in orientifold compactifications of Type IIB
superstring theory, and their no-scale structure gives rise to the
interesting possibility of having a dynamically generated
hierarchy of scales below the Planck scale \cite{gkp} -
\cite{kst}. In the present work  we show that all these
spontaneously broken models can  be viewed as non semisimple
gauging of ``flat groups" in extended supergravity, originally
considered in the generalized dimensional reduction of
Scherk--Schwarz \cite{ss}.

We will mainly consider the example of $N=8$  gauged supergravity
in four dimensions. The existence of a $\fso(1,1)$ grading  in the
Lie  algebra of the duality group $\rE_{7,7}
$ will turn out to be
essential. This grading comes from the decomposition
\begin{equation}\fe_{7,7}= \fe_{6,6}+\fso(1,1)+ \fp,\qquad
\fp=\mathbf{27_{-2}}+\mathbf{27'_{+2}},\label{e77}\end{equation}
where $\fp$ carries the mentioned representations of
$\fe_{6,6}+\fso(1,1)$.

Gauged supergravities based on the above decomposition originate
models that differ from previously constructed gaugings \cite{dwn}
- \cite{hw} and were overlooked in previous classifications
\cite{cfgtt}. They allow a construction of the Scherk--Schwarz
models from a purely four dimensional perspective.

Although in $N=8$ supergravity the structure of $\rE_{7,7}$ is
essential we will show that in a large class of models the
manifold of the scalars has  isometries which include $\fso(1,1)$
and translations. The $\fso(1,1)$ provides a  grading of the
isometry algebra as well as of the symplectic space of electric
and magnetic field strengths. For example, all $N=2$
supergravities based on special geometry with a cubic prepotential
have this property \cite{wlp,ckpdfwg}. It is then not surprise
that all these models have a five dimensional origin \cite{gst}.

The flat group whose gauging corresponds to the Scherk--Schwarz
mechanism has the following structure,
\begin{eqnarray}&&[X_\Lambda, X_0]=f_{\Lambda 0}^\Sigma
X_\Sigma,\\&&[X_\Lambda,X_\Sigma]=0\qquad \Lambda=1,\dots
27,\label{cr}\end{eqnarray} where $X_\Lambda$ is in the
$\mathbf{27'_{+2}}$ in (\ref{e77}) and $X_0$ is a generic Cartan
generator of $\fusp(8)\subset \fe_{7,7}$. The gauge group is then
a semidirect product of two abelian factors. Note that this group
is not a subgroup of any previously considered electric subgroup
of $\rE_{7,7}$. Flat groups of this kind exist in all four
dimensional extended supergravity models which have a five
dimensional origin, and in particular, in many no-scale
supergravity models obtained by turning on brane fluxes. Partial
supersymmetry breaking of these models correspond to different
Higgs branches of  the flat groups.

The paper is organized as follows. In section \ref{electric} we
discuss non isomorphic electric subgroups of $\rE_{7,7}$ and show
that a choice which is not a subgroup of $\rSL(8,\R)$ gives rise
to the gauging of the Scherk--Schwarz spontaneously broken
supergravity. In section \ref{lagrangian} we give the relation
between the four dimensional $N=8$ Lagrangian in standard form
with the one obtained by Sezgin and van Nieuwenhuizen \cite{sn} by
generalized dimensional reduction. In particular, we give the
relevant terms in the bosonic sector which are due to the gauging
of a flat group. Interestingly enough, these terms include
unconventional Chern-Simons terms for vector fields, due to the
gauging of translational isometries (Peccei-Quinn symmetries) and
to the fact that the scalar axions are charged under the gauge
group. In section \ref{lower} we discuss theories with lower
supersymmetries in the perspective of the gauging of flat groups.

\section{\label{electric}Electric subgroups of $N=8$ supergravity}

In $N=8$, $d=4$ supergravity there are 28 vector gauge potentials
$Z_\mu^\Lambda$. Their field strengths $F^\Lambda$, together with
their magnetic duals $G_\Lambda=  \frac 12\frac{\partial
\mathcal{L}}{\partial \, {{^*\!F}^\Lambda}}$\footnote{We have used
the following normalization for the kinetic Lagrangian of the
vectors: $${\mathcal L}= \Re ({\mathcal N}_{\Lambda\Sigma})
F^{\Lambda}_{\mu\nu } {^*\!F}^{\Sigma \mu\nu}+\Im ({\mathcal
N}_{\Lambda\Sigma}) F^{\Lambda}_{\mu\nu}F^{\Sigma\mu\nu} \mbox{,
with } {^*\!F}^{\mu\nu}\equiv \frac 12 \epsilon^{\mu\nu\rho\sigma}
F_{\rho \sigma}. $$}
 transform in the (56
dimensional) fundamental representation of $\rE_{7,7}$.
$\rE_{7,7}$ is embedded in $\rSp(56,\R)$ via the fundamental
representation, which is then said to be a symplectic
representation. This implies that the matrices of  the Lie
algebra $\fe(7,7)$   are of the form
\begin{equation}\begin{pmatrix}a & b\\c&-a^T\end{pmatrix},\qquad
b=b^T,\; c=c^T,\label{symp}\end{equation} with $a,b, c, d$ being
$28\times 28$ matrices. A generator of the symplectic algebra
$\fsp(56)$ has $a, b, c$ arbitrary ($b,c$ symmetric)  but the
generators belonging to the subalgebra $\fe_{7,7}$ have some
restrictions on these entries.

An electric subgroup of $\rE_{7,7}$ is any subgroup with $b=0$.
Such a subgroup acts linearly on the 28 dimensional space of the
electric field strengths (or vector potentials). If the 28 gauge
potentials are in the adjoint representation of an electric
subgroup, then it is in principle possible to gauge it.

The standard example is $\rSO(8)$ gauged supergravity \cite{wn}.
$\rSO(8)$ is  the  maximal compact subgroup of $\rSL(8,\R)$, which
is a maximal electric subgroup of $\rE_{7,7}$. In the
representation (\ref{symp}) it has not only $b=0$ but also $c=0$.
The reason is that
$\mathbf{56}\rightarrow\mathbf{28}+\mathbf{28'}$ \footnote{We will
denote by $\mathbf{r'}$ the contragradient representation of
$\mathbf{r}$. } under $\rSL(8,\R)$, so the field strengths are not
mixed with their duals by an $\rSL(8,\R)$ transformation.

We would like to give a parametrization of the group $\rE_{7,7}$
which depicts the embedding of different electric subgroups. If we
think about $N=8$ supergravity in $d=4$ as obtained by dimensional
reduction of $N=8$ supergravity in $d=5$, it is pretty obvious
that it must be possible to choose an electric subgroup of
$\rE_{7,7}$ which contains $\rE_{6,6}$. Indeed, $\rE_{6,6}$ has a
linear action on the 27 vector potentials in dimension five, so it
will have  it also on the dimensionally reduced vectors. The
28$^{\rm th}$ four dimensional vector comes from the metric, and
it  is an  $\rE_{6,6}$ singlet. This suggests that we should look
for new electric subgroups of $\rE_{7,7}$ by considering the
decomposition of the representations of $\rE_{7,7}$ under the
subgroup $\rE_{6,6}\times \rSO(1,1)$ (this subgroup is maximal as
a reductive subgroup of $\rE_{7,7}$, but it is not a maximal
subgroup, as we will see later).
 The  fundamental representation  decomposes as follows
 (the subindex indicates the charge under SO(1,1))
 $$\begin{CD}\mathbf{56}@>>\rE_{6,6}\times \rSO(1,1)>\mathbf{27_{+1}}+
 \mathbf{27'_{-1}}+\mathbf{1_{+3}}+\mathbf{1_{-3}},\end{CD}$$ with
 $\mathbf{1_{+3}}$ being the new vector that comes from the metric
 when performing the dimensional reduction. The adjoint
 representation decomposes as
\begin{equation}\begin{CD}\mathbf{133}@>>\rE_{6,6}\times
\rSO(1,1)>\mathbf{78_{0}}+
 \mathbf{1_{0}}+\mathbf{27_{-2}}+\mathbf{27'_{+2}};
 \end{CD}\label{adj}\end{equation}
$\mathbf{78_{0}}$ is  the adjoint of $\rE_{6,6}$, $\mathbf{1_{0}}$
is the generator of the rescaling of the radius of $S^1$, the
$\mathbf{27'_{+2}}$ are the shift symmetries acting on the axions
coming from the fifth component  of the 27 five dimensional
vectors, $\mathbf{27_{-2}}$ are the additional transformations,
not implementable on the vector potentials, which complete the
algebra of $\rE_{7,7}$.
\par
>From (\ref{adj}) it follows that $\fe_{7,7}$ has a grading under
$\fso(1,1)$, $$\fe_{7,7}=\fl_0+\fl_{+2}+\fl_{-2}.$$ The
representation {\bf 56} is a graded representation. The underlying
vector space is decomposed as  $V=V^+\oplus V^-$, this
decomposition being compatible  with the grading of the Lie
algebra. $V^+$ is the space of the electric field strengths (and
potentials) and $V^-$ the space of their magnetic duals.  In
particular we have that
\begin{eqnarray*}
&X\in \fl_0, \quad &X:V^+\longrightarrow V^+,\\ &X\in \fl_{+2},
\quad &X:V^+\longrightarrow V^+\end{eqnarray*} so the non
semisimple subalgebra $\fl_0+\fl_{+2}$ has a linear action on
$V^+$. Therefore, the matrices of $\fl_0+\fl_{+2}$ in the
fundamental representation have $b=0$ (\ref{symp}). $\fl_0$ has
also $c=0$, while $\fl_{+2}$ has $a\neq 0, \; c\neq 0$. We write
the action of $\fl_{\pm 2}$ on $V^+\oplus V^-$:
\begin{equation}
\delta\begin{pmatrix}F^\Lambda \\ F \\ G_\Lambda \\
G\end{pmatrix}=
\begin{pmatrix} 0^\Lambda_{\ \Sigma} & -{t'}^\Lambda & d^{\Lambda\Sigma\Gamma}t_\Gamma &
0^\Lambda\\ - t_\Sigma & 0  & 0_\Sigma &0\\
 d_{\Lambda\Sigma\Gamma}{t'}^\Gamma &0_\Lambda& 0_\Lambda^{\ \Sigma}& t_\Lambda \\
0_\Sigma & 0& {t'}^\Sigma &0\end{pmatrix} \begin{pmatrix}F^\Sigma
\\ F
\\ G_\Sigma
\\ G\end{pmatrix}.\label{gaugegroup}
\end{equation}
$t_\Lambda$ and ${t'}^\Lambda$ are the parameters of the
transformation and $d_{\Lambda\Sigma\Gamma}$ is the  symmetric
invariant tensor  of the representation $\mathbf{27}$ of
$\rE_{6,6}$ ($d^{\Lambda\Sigma\Gamma}$ of the $\mathbf{27'}$). The
matrices of $\fl_{-2}$ have $c=0$, $a\neq 0, b\neq 0$. We denote
the vector potentials as $( Z_\mu^\Lambda , Z_\mu^0=B_\mu)$, with
$\Lambda=1,\dots 27$. They transform as $V^+$, so we have
\begin{eqnarray*}&\delta_{\mathbf{27'_{+2}}} Z_\mu^\Lambda={t'}^\Lambda B_\mu\quad
&\delta_{ \mathbf{1_{0}}} Z_\mu^\Lambda=\lambda Z_\mu^\Lambda\\
&\delta_{\mathbf{27'_{+2}}} B_\mu=0\quad &\delta_{
\mathbf{1_{0}}}B_\mu=3\lambda B_\mu.\end{eqnarray*} The above
transformation properties under translations and $\rSO(1,1)$ are
common to a large class of supergravity models with different
number of supersymmetries.

Finally, let us discuss the scalar sector.
 The  coset of the scalars in  four and five dimensions are respectively
 $\rE_{7,7}/\rSU(8)$ and  $\rE_{6,6}/\rUSp(8)$. The corresponding  Cartan
 decompositions are
\begin{eqnarray*}&\fe_{7,7}=\fsu(8)+\fp,\qquad &\fp=\mathbf{70}\;
 {\rm of}\; \rSU(8)\\&\fe_{6,6}=\fusp(8)+\fp'\qquad  &\fp=\mathbf{42}\;
 {\rm of}\; \rUSp(8),\end{eqnarray*}
 and we have that the $\mathbf{70}$ of $\rSU(8)$ is decomposed
$$\begin{CD}\mathbf{70}@>>\rUSp(8)>\mathbf{42}+\mathbf{27}+\mathbf{1}.
\end{CD}$$
The physical meaning of the above decomposition is that the
scalars in the $\mathbf{27}$ come from the fifth component of the
27 five dimensional vectors and the singlet from the $g_{55}$
component of the metric (radius of $S^1$).

\section{\label{lagrangian} Standard form of $N=8$ Scherk--Schwarz supergravity and gauging of flat groups}
We want to compare the maximally extended ($N=8$) supergravity in
four dimensions with the theory found by Sezgin and Van
Nieuwenhuizen \cite{sn} through the Scherk--Schwarz dimensional
reduction from five dimensions.

Let us first consider the case where all the mass parameters are
set to zero (standard dimensional reduction).  In dimension five
the U-duality group is $\rE_{6,6}$, and it acts linearly on the 27
vector potentials $\hat A_{\hat\mu}^\Lambda, $ with $
\hat\mu=1,\dots 5$ and $\Lambda=1,\dots 27$.  We will denote the
quantities in five dimensions with a hat, ``~$\hat{\hbox{ }}$~",
to distinguish them from the four dimensional ones. The local
symmetries acting on these vector potentials are general
coordinate transformations in five dimensions with parameters
$\hat\xi^{\hat\mu}(x^{\hat\nu})$ and 27 abelian U(1) gauge
transformations $\Xi^\Lambda(x^{\hat\nu})$. When performing the
reduction to four dimensions, the local symmetries that remain
are:  four dimensional general coordinate transformations with
parameters $\xi^\mu(x^\nu)$ with $\mu,\nu=1,\dots 4$, a gauge
transformation with parameter $\xi^5(x^\nu)$, and the U(1) gauge
transformations with parameters $\Xi^\Lambda(x^{\nu})$. Explicitly
these transformations read
\begin{eqnarray*}
&&\delta_{\xi^5}A_\mu^\Lambda= \partial_\mu\xi^5 A_5^\Lambda\\
&&\delta_\Xi
A_\mu^\Lambda=\partial_\mu\Xi^\Lambda\\&&\delta_{\xi^5}A^\Lambda_5=0\\&&\delta_{\Xi}
A_5^\Lambda=0,
\end{eqnarray*}
 We will denote $A_5^\Lambda=a^\Lambda$. There is also a global  ($x$-independent) invariance
$a^\Lambda\rightarrow a^\Lambda+{t'}^\Lambda$.  We denote by
$B_\mu$ and $\phi$  the vector and scalar coming from the
reduction of the vielbein, $$\hat V_{\hat\mu}^{\hat a}=(V_\mu^a,
V_\mu^5=V^5_5B_\mu, V^5_5=e^{2\phi})$$ with transformation
$$\delta_{\xi^5}B_\mu=\partial_\mu\xi^5.$$ The combinations
$$Z_\mu^\Lambda=A_\mu^\Lambda-a^\Lambda B_\mu$$ are inert under
$\xi^5$ and so $Z^\Lambda_\mu$ and $B_\mu$ are genuine four
dimensional gauge fields. Note, however, that under the global
translation ${t'}^\Lambda$, $Z^\Lambda$ transforms $$\delta
Z_\mu^\Lambda=-{t'}^\Lambda B_\mu.$$ The meaning of this
transformation is that the 28 vectors $(Z_\mu^\Lambda,  B_\mu)$
form a 28 dimensional indecomposable representation of the 27
dimensional translation group (${t'}^\Lambda$). The action given
in Ref.\cite{sn} is also invariant under the following $\rSO(1,1)$
transformation (with parameter $\lambda$)
\begin{eqnarray*}
&&\phi'=\phi-\lambda\\&&{Z'}^\Lambda_\mu=e^\lambda
Z_\mu^\Lambda\\&&{B'}_\mu
=e^{3\lambda}B_\mu\\&&{a'}^\Lambda=e^{-2\lambda}a^\Lambda.\end{eqnarray*}
We observe  that the group generated by $\rSO(1,1)$ and the 27
global translations is precisely the same as the one discussed in
Section \ref{electric}, which appears in the decomposition of
$\rE_{7,7}$ under $\rE_{6,6}$.

In terms of the fields $\phi, a^\Lambda, B_\mu, Z_\mu^\Lambda$,
the Lagrangian (4.33) of \cite{sn} at zero masses  reduces  to the
following standard expression:\footnote{Note that our definitions
slightly differ from the ones in \cite{sn}. In particular, we have
defined the field strengths of the vectors as:
$Z^\Lambda_{\mu\nu}= \frac 12(\partial_{\mu}
Z^\Lambda_{\nu}-\partial_{\nu} Z^\Lambda_{\mu})$,
$B_{\mu\nu}=\frac 12 (\partial_{\mu} B_{\nu}-\partial_{\nu}
B_{\mu})$. Moreover, with respect to \cite{sn} we have redefined
$\frac \phi{\sqrt{3}} \rightarrow \phi$ and
$2d_{\Lambda\Sigma\Gamma}\rightarrow d_{\Lambda\Sigma\Gamma}$.}
\begin{eqnarray}
\mathcal{L}^{\mathrm{bos}}_{d=4} &=& -\frac 14 V R + \frac{3}{2} V
\partial_\mu \phi \partial^\mu \phi + \frac 14 V e^{-4\phi}
\hat{{\mathcal{N}}}_{\Lambda\Sigma} \partial_\mu a^\Lambda
\partial^\mu a^\Sigma+ \frac 1{24}VP_\mu^{abcd} P_{abcd}^\mu + \nonumber\\
&+& V{\Im}({\mathcal{N}}_{00}) B_{\mu\nu}B^{\mu\nu} +
2V{\Im}({\mathcal{N}}_{0\Lambda})Z^\Lambda_{\mu\nu}B^{\mu\nu}
+V{\Im}(\mathcal{
N}_{\Lambda\Sigma})Z^\Lambda_{\mu\nu}Z^{\Sigma\mu\nu} +
\nonumber\\
&+ &\frac 12\;\epsilon^{\mu\nu\rho\sigma}\left[{\Re}(\mathcal{
N}_{00}) B_{\mu\nu}B_{\rho\sigma} + 2{\Re}\mathcal ({N}_{\Lambda 0
})B_{\mu\nu}Z^\Lambda_{\rho\sigma} +{\Re}(\mathcal{
N}_{\Lambda\Sigma})Z^\Lambda_{\mu\nu}Z^{\Sigma}_{\rho\sigma}\right]\label{nglag}
\end{eqnarray}
in terms of the 28$\times$28 complex symmetric matrix
\begin{eqnarray*}
{\mathcal{N}}_{00}&=&\frac 1{3}d_{\Lambda\Sigma\Gamma} a^\Lambda
a^\Sigma a^\Gamma -\frac{\rm i}{2} \left(e^{2\phi} a^\Lambda
a^\Sigma \hat{\mathcal{N}}_{\Lambda\Sigma} + \frac 12 e^{ 6\phi}
\right)\\
{\mathcal{N}}_{\Lambda 0}&=& \frac 1{2}d_{\Lambda\Sigma\Gamma}
a^\Sigma a^\Gamma-\frac{\rm i}{2}
e^{2\phi}\hat{\mathcal{N}}_{\Lambda\Sigma}a^\Sigma\\ \mathcal
{N}_{\Lambda\Sigma}&=& d_{\Lambda\Sigma\Gamma} a^\Gamma
-\frac{\rm i}{2} e^{2\phi}\hat{\mathcal{N}}_{\Lambda\Sigma}.
\end{eqnarray*}
$P_\mu^{abcd}$ is the $\rE_{6,6}/\rUSp(8)$ vielbein, $V=\det
V_\mu^a$ and $\hat {\mathcal{N}}_{\Lambda\Sigma}$ is the five
dimensional ($\rSO(1,1)$ invariant) vector kinetic matrix.

The representation of $\rE_{7,7}$ on the 56 dimensional vector
space of the field strengths and their duals is symplectic. For
the moment being, let ${\mathcal{F}}$ denote the vector
$(F^\Lambda, F)$ and ${\mathcal{G}}$ $(G_\Lambda, G)$.  In terms
of the self dual combinations ${{\mathcal{F}}^+}=\frac 12
\left({\mathcal{F}}+{\rm i}\,^*\!{\mathcal{F}}\right)$,
${{\mathcal{G}}^+}=\frac 12\left({\mathcal{G}} +{\rm
i}\,^*\!{\mathcal{G}}\right)$, a generic element of the symplectic
 group would act as $$\begin{pmatrix} {{\mathcal{F'}}}^+\\
{{\mathcal{G'}}}^+\end{pmatrix}=\begin{pmatrix}A&B\\C&D\end{pmatrix}\begin{pmatrix}
{\mathcal{F}}^+\\ {\mathcal{G}}^+\end{pmatrix},\quad A^TC\; {\rm
and }\; B^TD\;{\rm symmetric}, \quad A^TD-C^TB=\id.$$ Then the
matrix ${\mathcal{N}}$ transforms as
$${\mathcal{N}}'={(C+D{\mathcal{N}})}{(A+B{\mathcal{N}})^{-1}}.$$
The fractional formula allows us to compute the non linear
transformations of the fields under the upper block triangular
transformation corresponding to the ${\bf 27_{-2}}$ elements of
$\fe_{7,7}$  in (\ref{gaugegroup}),
\begin{eqnarray*}
&&\delta{\mathcal{N}}_{\Lambda\Sigma}=-{\mathcal{N}}_{\Lambda\Pi}{\mathcal{N}}_{\Delta\Sigma}
d^{\Pi\Delta\Gamma}{t}_\Gamma + t_\Lambda{\mathcal{N}}_{\Sigma
0}+t_\Sigma{\mathcal{N}}_{\Lambda0}\\
&&\delta{\mathcal{N}}_{\Lambda
0}=-{\mathcal{N}}_{\Lambda\Sigma}{\mathcal{N}}_{\Delta 0
}d^{\Sigma\Delta\Gamma}{t}_\Gamma
+t_\Lambda{\mathcal{N}}_{00}\\&&\delta{\mathcal{N}}_{00}=-{\mathcal{N}}_{\Lambda
0 }{\mathcal{N}}_{\Gamma 0
}d^{\Lambda\Gamma\Delta}{t}_\Delta\end{eqnarray*} The
infinitesimal transformation of ${\mathcal{N}}$ under the
translational symmetries $\mathbf{27'_{+2}}$ (lower block
triangular) in (\ref{gaugegroup}) is
\begin{eqnarray*} &&\delta{\mathcal{N}}_{\Lambda\Sigma}=d_{\Lambda\Sigma\Delta}{t'}^\Delta\\
&&\delta{\mathcal{N}}_{\Lambda
0}={\mathcal{N}}_{\Lambda\Sigma}{t'}^\Sigma\\&&\delta{\mathcal{N}}_{00}=2{\mathcal{N}}_{\Lambda
0 }{t'}^\Lambda\end{eqnarray*}

In particular, we note that only
$\Re({\mathcal{N}}_{\Lambda\Sigma})$ (theta term) transforms non
linearly under the translations $t'$.

\bigskip

We consider now the semidirect product of the 27 translations with
parameters ${t'}^\Lambda$ with a generic element of the Cartan
subalgebra of $\rUSp(8)$, maximal compact subgroup of $\rE_{6,6}$.
Its Lie algebra  is of the form  (\ref{cr}) with $f_{\Lambda
0}^\Sigma$ the matrix of the  Cartan element in the representation
$\mathbf{27}$ of $\rUSp(8)$. We will denote it by
$M_\Lambda^\Sigma$.  Since $\rUSp(8)$ has rank four,
$M_\Lambda^\Sigma$ depends on four parameters, $m_i, \; i=1,\dots
4$ and its 27 eigenvalues are functions of these four paramenters.
They  are given by \cite{ss} $$\pm{\rm i}(m_i\pm m_j), \; i<j$$
 and 3 eigenvalues are 0 \footnote{The physical masses of the vector bosons actually are
proportional to the above eigenvalues with a moduli dependent
prefactor $e^{-3\phi}$, as it can be seen from inspection of the
Lagrangian (\ref{nglag}). The same observation holds also for the
masses of the other particles, as given in \cite{ss,css} }. This
means that there are 3 linear
 combinations of $X_\Lambda$ which actually commute with $X_0$.

 We choose
this group to perform the gauging of $N=8$ supergravity.
$Z_\mu^\Lambda$ are the gauge connections for $X_\Lambda$
generators and $B_\mu$ is the U(1) gauge connection. It is
straightforward  to see that the gauge transformations of the
connection fields
 are as follows
 \begin{eqnarray*}
 &&\delta Z_\mu^\Lambda=\partial_\mu\Xi^\Lambda
 +\Xi^0M_\Sigma^\Lambda Z^\Sigma_\mu-\Xi^\Sigma M_\Sigma^\Lambda B_\mu\\
 &&\delta B_\mu=\partial_\mu\Xi^0.\end{eqnarray*}
 The field strengths are
 \begin{eqnarray*}&& F_{\mu\nu}^\Lambda=\frac 12\left(\partial_{\mu}
 Z_{\nu}^\Lambda-\partial_{\nu}
 Z_{\mu}^\Lambda-M_\Sigma^\Lambda (Z^\Sigma_{\nu} B_{\mu}-Z^\Sigma_{\mu} B_{\nu})\right)\\
 &&B_{\mu\nu}=\frac 12\left(\partial_\mu B_\nu-\partial_\nu B_\mu\right).\end{eqnarray*} The
 gauge transformation of the axion fields $a^\Lambda$ is
 $$\delta a^\Lambda=M^\Lambda_\Sigma\Xi^\Sigma+\Xi^0M^\Lambda_\Sigma
 a^\Sigma,$$ and their covariant derivatives
 $$\nabla_\mu a^\Lambda=\partial_\mu  a^\Lambda-M^\Lambda_\Sigma
 a^\Sigma B_\mu - M_\Sigma^\Lambda Z_\mu^\Sigma.$$

 In order to write the gauge completion of the Lagrangian
 (\ref{nglag}), we first observe that
 $\hat{\mathcal{N}}_{\Lambda\Sigma}$ is invariant under the gauge
 transformations $\Xi^\Lambda$, but transforms under $\Xi^0$ as
 follows
 $$\delta\hat{\mathcal{N}}_{\Lambda\Sigma}=\Xi^0M_\Lambda^\Delta\hat
 {\mathcal{N}}_{\Sigma\Delta}+ (\Lambda\leftrightarrow\Sigma).$$
 It then follows that under $\Xi^0$
 \begin{eqnarray*}
 &&\delta{\mathcal{N}}_{\Lambda\Sigma}=\Xi^0M_\Lambda^\Delta{\mathcal{N}}_{\Sigma\Delta}+(\Lambda\leftrightarrow\Sigma)\\
&&\delta{\mathcal{N}}_{\Lambda
0}=\Xi^0M_\Lambda^\Delta{\mathcal{N}}_{\Sigma 0}\\
&&\delta{\mathcal{N}}_{00}=0,\end{eqnarray*} and under
$\Xi^\Lambda$
\begin{eqnarray*}
 &&\delta{\mathcal{N}}_{\Lambda\Sigma}=d_{\Lambda\Sigma\Delta}M_\Pi^\Delta\Xi^\Pi\\
&&\delta{\mathcal{N}}_{\Lambda
0}={\mathcal{N}}_{\Lambda\Sigma}M_\Pi^\Sigma\Xi^\Pi\\
&&\delta{\mathcal{N}}_{00}=2{\mathcal{N}}_{\Lambda
0}M_\Pi^\Lambda\Xi^\Pi.\end{eqnarray*}
 The gauge completion of the
lagrangian (\ref{nglag}) is then obtained by replacing
\begin{eqnarray*}&&F_{\mu\nu}^\Lambda\mapsto F_{\mu\nu}^\Lambda
\;\; {\rm non \;abelian}\\ &&\partial_\mu a^\Lambda\mapsto
\nabla_\mu a^\Lambda\\&&P_\mu^{abcd}\mapsto
P_\mu^{abcd}-P_5^{abcd}B_\mu,\end{eqnarray*} where $P_5^{abcd}$ is
defined in (4.3) and (4.33) of Ref. \cite{sn} and has the property
that $$\delta_{\Xi^0}P_\mu^{abcd}=P_5^{abcd}\partial_\mu\Xi^0,$$
and adding the extra term \begin{equation} \mathcal{L}^{\rm
extra}=\frac
13d_{\Lambda\Sigma\Pi}M^\Pi_\Delta\epsilon^{\mu\nu\rho\sigma}Z_\mu^\Lambda
Z_\nu^\Delta(\partial_\rho Z_\sigma^\Sigma-\frac 34M_\Gamma^\Sigma
Z_\rho^\Gamma B_\sigma). \label{lextra} \end{equation}
 Note that the second term in (\ref{lextra}) is in fact
 identically zero here, due to the symmetry property of
 $d_{\Lambda\Sigma\Pi}$.

In the Scherk--Schwarz theory, the term (\ref{lextra}) comes from
the generalized dimensional reduction of the Chern-Simons term
\cite{sn}, but in a four dimensional setting it is required
because of the non linear transformation
$$\delta\Re({\mathcal{N}}_{\Lambda\Sigma})=d_{\Lambda\Sigma\Pi}M^\Pi_\Gamma\Xi^\Gamma,
$$ as shown in  (3.16) of Ref. \cite{wlp}.

Here $d_{\Lambda\Sigma\Pi} M^\Pi_\Gamma =C_{\Gamma
,\Lambda\Sigma}$ of \cite{wlp} and it satisfies the property
$$C_{\Gamma ,\Lambda\Sigma} +C_{\Lambda
,\Sigma\Gamma}+C_{\Sigma,\Gamma \Lambda}=0$$ as a consequence of
the fact that $d_{\Lambda\Sigma\Pi}$ is an invariant tensor of
$E_{6(6)}$.

Supersymmetry also requires a scalar potential $$U= \frac
1{24}e^{-6\phi}P_{5{abcd}}P_{5}^{abcd}$$ as shown in Ref.
\cite{sn}. We do not discuss here the fermionic sector of the
theory whose gauge completion can also be obtained in a standard
manner.

\section{\label{lower} Gauging of flat groups in $N=2$ four dimensional supergravities}
We want to extend the gauging of flat groups to a class of $N=2$
four dimensional supergravities which have a five dimensional
interpretation. The graviton multiplet has fields
$$(V_\mu^a,\psi_{\mu A}, Z^0_\mu).$$ It is well known that the
interactions of $N=2$ vector multiplets with fields
$$(Z_\mu^\Lambda, \lambda_A^\Lambda, z^\Lambda), \qquad
\Lambda=1,\dots n_v,$$ are described by the special geometry of
the scalar sigma model. The special geometry \cite{wlp} of the
Kaehler manifold of the scalar fields is completely specified by
an holomorphic prepotential of $n_v+1$ variables, homogeneous of
degree two  $$\mathcal{F}(x^\Lambda,
x^0)={x^0}^2{f}(\frac{x^\Lambda}{x^0}),$$ where in special
coordinates,  $$z^\Lambda=\frac{x^\Lambda}{x^0}.$$ The holomorphic
functions $$(x^\Lambda, x^0, F_\Lambda=\frac{\partial
\mathcal{F}}{\partial x^\Lambda},F_0=\frac{\partial
\mathcal{F}}{\partial x^0})$$ are a local section of a flat
symplectic bundle (with structure group $\rSp(2n_v+2,\R)$) over
the special Kaehler manifold \cite{st,cdf}.
 The duality group $G$ is a subgroup of $\rSp(2n_v+2,\R)$. The representation
  of $G$ acting on the field strengths $({F^+}^\Lambda_{\mu\nu},
{{F^+}^0}_{\mu\nu})$ and their duals $({G^+}_{\Lambda\mu\nu},
{G^+}_{0\mu\nu})$ is a symplectic representation.

If we choose a cubic prepotential of the form
$$\mathcal{F}=\frac1{3!}\frac{c_{\Lambda\Sigma\Delta}x^\Lambda
x^\Sigma x^\Delta}{x^0},$$ then the Kaehler manifold has always
$n_v+1$ isometries which form a group
$\rSO(1,1)\circledS\,\R^{n_v}$ (semidirect product) acting on the
coordinates $x$ as follows
\begin{eqnarray*}&\delta x^\Lambda=\lambda x^\Lambda,\qquad &\delta
x^0=3\lambda x^0\\&\delta x^\Lambda={t'}^\Lambda x^0\qquad &\delta
x^0=0.\end{eqnarray*} This transformation induces a symplectic
action on the symplectic sections and the electromagnetic field
strengths as follows
$$\begin{pmatrix}\lambda&{t'}^\Lambda&0&0\\0&3\lambda&0&0\\
c_{\Lambda\Sigma\Delta}{t'}^\Delta&0&-\lambda&0\\0&0&-{t'}^\Lambda&-3\lambda\end{pmatrix}.$$

The lower block triangular transformations have the same form of
the $\fl_0+\fl_+$ of $\fe_{7,7}$. This should not come as a
surprise since all these theories can be obtained by dimensional
reduction of $N=2$ supergravity in five dimensions with $n_v-1$
vector multiplets  whose real special geometry is defined by the
cubic surface \cite{gst} $$c_{\Lambda\Sigma\Delta}y^\Lambda
y^\Sigma y^\Delta=1.$$ If the real special geometry in five
dimensions  has no isometries ($c_{\Lambda\Sigma\Delta}$ are
generic), and in the absence of hypermultiplets, the only global
symmetry in five dimensions is the SU(2) R-symmetry and the
Scherk--Schwarz generalized dimensional reduction would correspond
to the gauging of the U(1) Cartan element of SU(2), giving mass
only to the gravitinos and the gauginos, which are the only fields
charged under this U(1). Supersymmetry would be broken while the
$\rU(1)\circledS \, \R^{n_v}$ group would remain unbroken.

 In the case where there is a group of isometries $G'$ acting on
 the real special manifold (for dimension 5), then we could use
 for the Scherk--Schwarz mechanism a U(1) which has also a component
 on the Cartan element of the maximal compact subgroup of $G'$.
 A non trivial flat group would emerge in four dimensions whose
 gauging would be similar to the gauging of $N=8$
 supergravity discussed in section \ref{electric} .

 As an illustration, let us consider models where the
 real special geometry is a coset space $G'/H'$ and the $n_v$ five
 dimensional vectors belong to a linear representation $R(n_v)$ of $G'$.
 Kaluza-Klein dimensional reduction implies that the U-duality group in four dimensions $G$ has an
 $\rSO(1,1)$ grading when decomposing with respect to $G'\subset G$
 $$\fg=\fg'_0+ \mathbf{1_0}+ \fr(n_v)_{\mathbf{-2}} +
 \fr'(n_v)_{\mathbf{+2}}.$$
 ($\fr$ denotes the representation space of $R_{n_v}$). Moreover,
 the $2n_v+2$ dimensional symplectic representation  $R(2n_v+2)$ of the four
 dimensional duality group $G$ has the following decomposition
 under $G'\times \rSO(1,1)$
 $$\fr(2n_v+2)=\fr(n_v)_{\mathbf{+1}}+\mathbf{1_{+3}}
 +\fr'(n_v)_{\mathbf{-1}}+\mathbf{1_{-3}}.$$
 Note that these decompositions and the $\rSO(1,1)$ grading are
 universal and do not depend on the choice of $G'$. Indeed, this
 grading has a Kaluza-Klein origin.

 We give in Table \ref{examples} the examples corresponding to the
 exceptional five dimensional cosets. The flat group of dimension
 $n_v+1$ has structure constants which depend on ${\rm
dim(CSA)}_{H'}+1$ parameters.

\begin{table}[ht]
\begin{center}
\begin{tabular} {|c|c|c|c|c|}
\cline{1-5}  $G'/H'$&$G/H$& $\fr(n_v)$ &$\fr(2n_v+2)$&${\rm
dim(CSA)}_{H'}$
\\ \cline{1-5}$\rSL(3,\R)/\rSO(3)$&$\rSp(6,\R)/\rU(3)$&{\bf
6}&{\bf 14}&1\\
$\rSL(3,\C)/\rSU(3)$&$\rU(3,3)/\rU(3)\times\rU(3)$&{\bf 9}&{\bf
20}&2\\ $\rSU^*(6)/\rUSp(6)$&$\rSO^*(12)/\rU(6)$&{\bf 15}&{\bf
32}&3\\ $\rE_{6,-26}/\rF_4$&$\rE_{7,-25}/\rE_6\times\rSO(2)$&{\bf
27}&{\bf 56}&4
\\\cline{1-5}
\end{tabular}
\caption{Exceptional $N=2$ supergravities.}\label{examples}
\end{center}
\end{table}

\bigskip

Most of the spontaneously broken models studied in Ref.
\cite{adfl} have a dynamical origin analogous to the one discussed
in the present investigation. However, since there are models
which break an odd number of supersymmetries, one can consider
gaugings  that do not have a five dimensional interpretation. This
should be the case for some models obtained by turning on brane
fluxes \cite{gkp,fp,kst}.

\section*{Acknowledgements}

S. F. would like to thank Raymond Stora for enlightening
conversations and the Dipartimento di Fisica, Politecnico di
Torino for its kind hospitality during the  completion of this
work. Work supported in part by the European Comunity's Human
Potential Program under contract HPRN-CT-2000-00131 Quantum
Space-Time, in which L. A., R. D. and M. A. Ll. are associated to
Torino University. The work of S. F. has also  been supported by
the D.O.E. grant DE-FG03-91ER40662, Task C.

\end{document}